\documentclass[oneocolumn]{aastex631}

\usepackage{rotating}

\usepackage{float}

\shorttitle{Dependence of stellar differential rotation}
\shortauthors{Araújo \& Valio}

\graphicspath{{./}{figures/}}

\begin{document}

\title{\Large Dependence of stellar differential rotation on effective temperature and rotation: an analysis from starspot transit mapping}

\correspondingauthor{Alexandre Araújo}

\email{adesouza.astro@gmail.com, avalio@craam.mackenzie.br}

\author{Alexandre Araújo}
\affiliation{Centro de Rádio Astronomia e Astrofísica Mackenzie, Universidade Presbiteriana Mackenzie\\
Rua da Consolação, 860, São Paulo, SP - Brazil}

\author{Adriana Valio}
\affiliation{Centro de Rádio Astronomia e Astrofísica Mackenzie, Universidade Presbiteriana Mackenzie\\
Rua da Consolação, 860, São Paulo, SP - Brazil}

\begin{abstract}
Stellar rotation is crucial for studying stellar evolution since it provides information about age, angular momentum transfer, and magnetic fields of stars. In the case of the Sun, due to its proximity,  detailed observation of sunspots at various latitudes and longitudes allows the precise estimate of the solar rotation period and its differential rotation. Here, we present for the first time an analysis of stellar differential rotation using starspot transit mapping as a means of detecting differential shear in solar-type and M stars. The aim of this study is to investigate the relationship between rotational shear, $\Delta\Omega$, with both the star’s effective temperature ($T_{\textrm{eff}}$) and average rotation period ($P_{\textrm{r}}$). We present differential rotation profiles derived from previously collected spot transit mapping data for 13 slowly rotating stars ($P_{\textrm{rot}} \ge$4.5 days), with spectral types ranging from M to F, which were observed by the Kepler and CoRoT satellites. Our findings reveal a significant negative correlation between rotational shear and the mean period of stellar rotation (correlation coefficient of -0.77), which may be an indicator of stellar age. On the other hand, a weak correlation was observed between differential rotation and the effective temperature of the stars. 
Overall, the study provides valuable insights into the complex relationship between stellar parameters and differential rotation, which may enhance our understanding of stellar evolution and magnetic dynamos.
\end{abstract}


\section{Introduction} \label{sec:intro}

Stellar evolution research heavily relies on studying the rotation of stars as it provides valuable insights into their age, transfer of angular momentum, and magnetic fields.
Stellar rotation, coupled with convective motions and strong magnetic fields within the stellar interior, produces various magnetic phenomena, including stellar spots, chromospheric plages, coronal loops, and flares \citep{berdy+05}. 
High-precision photometry from space missions has enabled systematic studies of stellar rotation, flares, and starspots, enhancing our understanding of stellar activity.
Especially, space mission data have revealed possibilities for understanding if activity phenomena on the Sun, such as sunspots, are similar to those observed on other solar-type stars.

Theoretical and observational studies on differential rotation are important for understanding the stellar dynamo. 
Differential rotation, similar to that of the Sun, is also observed in other stars \citep{reinhold2013rotation,mc+14}. 
However, the mechanisms influencing differential rotation in stars are still ambiguous due to conflicting results from different measurements in various time series \citep{reinhold2015rotation}. 


Based on the Sun, the differential latitudinal rotation of stars is usually described by a law of the form $\Omega (\alpha) = \Omega_{\textrm{e}} - \Delta \Omega\ \sin^{2} \alpha$, where $\alpha$ is the latitude, $ \Omega_\textrm{e}$ the rate of angular rotation at the equator, and $\Delta \Omega$ is the rotational shear, being the difference in the rate of rotation between the equator and the poles, $\Delta \Omega = \Omega_{\textrm{e}} - \Omega _{\textrm{p} }$. Another measured quantity is the relative differential rotation given by $\Delta \Omega/\bar \Omega$ (\%), where $\bar \Omega$ is the average rotation rate. For the Sun, $\Delta \Omega \simeq 0.05$ rd d$^{-1}$ \citep{Beck2000, Brajsa2001} and $\Delta \Omega/\bar \Omega \simeq $ 0.21 or 21\%.

Photometric modulations observed in stellar light curves, with periodicity corresponding to the stellar rotation, are typically associated with the presence of large spots that come into view and disappear as the star rotates. In the case of stars with transiting planets, the planet can pass in front of a starspot, resulting in a detectable signal within the transit light curve. This technique, known as spot transit mapping, enables the determination of physical properties of the spots, such as their radius, intensity, and location (longitude and latitude) \citep{silva+03}. By identifying the same spot in subsequent transits, the rotation period of the star can be estimated \citep[e.g.]{silva-valio2008}. 
Moreover, detecting starspots at different latitudes enables the precise estimate of differential rotation \citep{araujo+21}.

In this study, we analyze the stellar differential rotation estimated using the method of planetary transit spot mapping, which has been successfully applied to F-type \citep{valio2016starspots}, G-type \citep{valio+17,zaleski+19}, K-type \citep{araujo+21,valio22,zaleski2022dynamo}, and M-type stars \citep{zaleski2020,1651}. For the first time, we investigate correlations between stellar shear from transit mapping with the star's effective temperature and the average rotation period. In this study, we analyzed the rotational shear of 13 stars with transiting planets observed by the Kepler and CoRoT satellites. The following section provides an overview of the data and modeling methods used. In Section~\ref{sec:results}, we present a detailed analysis and discussion of our findings. Finally, we summarize the key finds and present our conclusions in Section~\ref{sec:conclusion}.

\section{Data and spot modeling} \label{sec:method}

The stars analyzed here were observed by the Kepler  \citep{borucki2010kepler} and CoRoT \citep{Baglin2006,auvergne2009corot} space missions. All thirteen stars host transiting exoplanets, enabling the detection of dark spots from their transit light curves. Table~\ref{tab:starparam} displays essential stellar parameters, including the star's name, mass (in solar mass), spectral type, stellar radius (in solar radius), and their respective references, in the columns from left to right. Moreover, the various characteristics of the transiting exoplanets are listed in Table~\ref{tab:planetparam}: name, radius in Jupiter radius and in degrees, orbital period (in days), semi-major axis (in AU), obliquity $\psi$, sky-projected value $\lambda$, and the star projected latitude of the exoplanet's transit (given by Eq.~\ref{eq:lat}), along with their corresponding references, in the columns from left to right.

\begin{table}
\setlength{\arrayrulewidth}{2\arrayrulewidth}
\centering
\small
\caption{Physical Parameters of Stars.}
\def\arraystretch{1.2}  
\setlength{\tabcolsep}{0.9mm}
\begin{tabular}{l|cccccccccc}
\hline 
\hline
Star & Mass[$M_{\odot}$]& Spectral Type & $R_{\textrm{star}}$[$R_{\odot}$] & Ref   \\
\hline
CoRoT-4  &1.16$^{+0.03}_{-0.02}$   & F8V  & 1.17$^{+0.01}_{-0.03}$ &     $(1)$\\
CoRoT-5 &1.00$\pm$0.02& F9V &1.186$\pm$0.040                            &     $ (1) $\\
CoRoT-6 &  1.05$\pm$0.05 &F9V&  1.025$\pm$0.026                        &     $(1)$\\
\hline
CoRoT-2 & 0.97$\pm$ 0.060 & G7V & 0.902$\pm$0.018                        &     $(1)$ \\
CoRoT-18  &   0.95$\pm$0.15 &  G9V &   1.00$\pm$0.13                     &     $(1)$\\
Kepler-17 & 1.16$\pm$0.060   & G2V & 1.05$\pm$0.03                     &     $(1)$    \\
Kepler-71 & 0.997$^{+0.03}_{-0.07}$   & G  & 0.89$\pm$0.05              &    $(2)$   \\
\hline 
CoRoT-8  & 0.880$\pm$0.040 & K1V  &0.770$\pm$0.020                       &    $(1)$ \\
Kepler-411 & 0.870$\pm$0.039  & K  & 0.820$\pm$0.018                     &    $(3)$  \\
Kepler-304  & 0.80$^{+0.020}_{-0.026}$   & K  & 0.658$^{+0.030}_{-0.028}$  &   $(4), (5)$   \\
Kepler-210 & 0.63  & K  & 0.69                                           &    $(6)$    \\
\hline
Kepler-45 & 0.59 $\pm$ 0.060   & M  & 0.55 $\pm$0.11                       &    $(1)$  \\
Kepler-1651A & 0.522$^{+0.033}_{-0.031}$ & M  & 0.503$^{+0.030}_{-0.027}$  &    $(7)$  \\
\hline
\end{tabular}
\label{tab:starparam}
\tablecomments{
(1) \citep{bonomo2017gaps};
(2) \citep{mathur2017revised};
(3) \citep{sun2019kepler}; 
(4) \citep{berger2018revised}; 
(5) \citep{morton2016false}; 
(6) \citep{ioannidis2014kepler};
(7) \citep{mann2017gold}. 
}
\end{table}

\begin{table*}
\setlength{\arrayrulewidth}{2\arrayrulewidth}
\centering
\small
\caption{Physical Parameters of Exoplanets.}
\def\arraystretch{1.2}  
\begin{tabular}{lllcllcllc}
\hline
\hline
Exoplanet & R$_{\textrm{p}}$[R$_{\textrm{J}}$] & R$_{\textrm{p}}$[$^\circ$]  &  $P_{\textrm{orb}}$[d] & $a$[AU] & $\Psi[^{\circ}]$ & $\lambda[^{\circ}]$ & lat[$^{\circ}$] & Ref \\
\hline
CoRoT-2 b        & 1.465  & 10.2     & 1.72   & 0.0281 & $<$ 4 & 4.7$\pm$12.3& -14.6      &1,2\\
CoRoT-4 b        & 1.190  & 6.1     & 9.20   & 0.09   &-&-& -1.74                        &1\\
CoRoT-5 b        & 1.330  & 6.7     & 4.03   & 0.049  &-&-& -27.2                        &1\\
CoRoT-6 b        & 1.166  & 7.0     & 8.88   & 0.085  &-&-& -16.4                        &1\\
CoRoT-8 b        & 0.562  & 5.9     & 6.21   & 0.063  &-&-& -29.4                        &1\\
CoRoT-18 b       & 1.310  & 7.5     & 1.90   & 0.0295 &20$\pm$20& -10$\pm$20  & -0.05    & 1,3\\
Kepler-17 b      & 1.312  & 8.0     & 1.48   & 0.0255 & $<10$&$<15$& -5.63                &1,4\\
Kepler-45 b      & 0.960  & 12.7     & 2.45   & 0.03   &$<11$&- & -33.9                   &5\\
Kepler-71 b      & 1.110  & 7.8     & 3.90   & 0.047  &$<6$&-& -5.40                     &6\\
Kepler-210 b     & 0.262  & 2.8      & 2.45   & 0.032  &-&-& -32.9                       &1,7\\
Kepler-411 b     & 0.214  & 1.4     & 3.00   & 0.0375 &-&-&-11.0                         &1\\
Kepler-411 c     & 0.394  & 2.6     & 7.83   & 0.0739 &-&-& -21.6                         &1\\
Kepler-304 b     & 0.273  & 2.6     & 3.30   &   0.0428    &-&-& -28.5                   & 1, 8\\ 
Kepler-1651 b    & 0.164  & 2.3     & 9.78   & 0.0619 &-&-& -32.2                        &1\\

\hline
\end{tabular}
\label{tab:planetparam}
\tablecomments{
1-exoplanet.eu;
2-\citep{alonso2008transiting};
3- \citep{hebrard2011transiting};
4- \citep{desert2011hot};
5- \citep{johnson2012characterizing};
6- \citep{howell2010kepler};
7- \citep{valio22};
8- \citep{zaleski2023}.
}
\end{table*}

\subsection{Starspot Transit Mapping}

The model used to analyze stellar spots is that presented in \cite{silva+03}, which was the first model to propose that an exoplanet could be used as a probe to detect the presence of spots on the surface of stars. This model allows for the physical characterization of spots in stars (spectral types FGK and M) observed by the CoRoT, Kepler, and TESS satellites, as well as future space missions.

The passage of the planet in front of the star can occult spots on the stellar surface, producing a slight increase in the luminosity during a short period (a few minutes) of the transit light curve, as described in \cite{silva+03}. This model, called $ECLIPSE$\footnote{Available as a Python code in https://github.com/Transit-Model-CRAAM/pipelineMCMC}, generates a synthesised two-dimensional limb darkened image of the star (according to given limb darkening coefficients), and assumes  that the planet is a dark disk with radius $\frac{R_{\textrm{\tiny p}}}{R_ {\textrm{\tiny star}}}$, where $R_{\textrm{\small p}}$ is the radius of the planet and $ R_{\textrm{\small star}}$ is the stellar radius. For each time interval, the position of the planet in its orbit is calculated according to its parameters: inclination angle $i$, semi-major axis $a$, and eccentricity $e$.
These input parameters as well as the planet radius and orbital period, plus the stellar limb darkening coefficients are initially obtained from the literature. These are then used as a first guess to a fit of the average of a few transit light curves that do not show spot signatures, from which  these parameters are further refined.

\subsection{Latitude and Longitudes of Starspots}

The  $ECLIPSE$ model \citep{silva+03} simulates the light curve of an exoplanet transit and allows the addition of circular dark (spots) or bright (faculae) features on the synthesised 2D image of the stellar disk. 
When the exoplanet occults a spot during its transit, a small increase in brightness is seen in the transit light curve, since the drop in total brightness is less than when the exoplanet transits a spotless region.

These small increases in the intensity of the light curve during spot occultation by the planet are seen as small “peaks”. The subtraction of a spotless star model from the light curve yields the residuals, where the “peaks” due to spots are enhanced and are more easily identified and modeled. The modeling of the excesses detected in the transit residuals yields the physical parameters of the spots such as size, intensity, and location (longitude and latitude) on the surface of the star. The topographic longitude of the spot, as seen from Earth, is obtained from the approximate time, $t_s$, with respect to transit centre time of the peak flux due to the spot in the residuals, given in hours, according to the following relationship:
\begin{equation}
    \textrm{lg}_{\textrm{topo}}=\arcsin\left[\frac{a\ \cos [90^{\circ}-(360^{\circ}{t_s}/24~ P_{\textrm{orb}})]}{\textrm{cos}(\textrm{lat})}\right],
    \label{eq:lg}
\end{equation}
\noindent where, $\textrm{lg}_{\textrm{topo}}$ is the longitude of the spot, $a$ is the semi-major axis, $P_{\textrm{orb}}$ is the orbital period of the planet, and $\textrm{lat}$ is the latitude of the transit projection onto the stellar disk given by:
\begin{equation}
    \textrm{lat} = \arcsin\left[\frac{a}{R_{\textrm{star}} \cos(i)}\right],
    \label{eq:lat}
\end{equation}
\noindent where $R_{\textrm{star}}$ is the radius of the star and $i$ the orbital inclination angle. The latitude band of the transit is defined by the planet size, which radius in degrees is listed in the third column of Table~\ref{tab:planetparam}.

From the intensity of the spots, its  temperature can be obtained by considering that both the stellar photosphere and the spot emit radiation like a blackbody (Eq.~2 of \cite{silva2010}).

Only spots detected within longitudes of $\pm$70$^{\circ}$ are considered in the model. This avoids the great uncertainties in  modeling of the part of the transit where there are large variations in brightness observed during the ingress and egress of the planet. 

\subsection{Rotation period at a given latitude}

The spot transit mapping model allows for high spatial resolution of spot's size and location on the stellar disk. When a star is transited by more than one planet, information about the spotted surface is obtained for more than one latitude \citep{araujo+21, valio22}, and the differential rotation is better constrained.
The same model presented in \cite{silva+03}, and described above, can be used to estimate the stellar rotation period at the transit latitude, by detecting the same spot group in a later transit \citep{silva2011}. 

To study the star's rotation profile, the spot observed longitude, $\textrm{lg}_{\textrm{topo}}$, needs to be converted to the reference frame that rotates with the star. For this, it is necessary to determine the rotational longitude, $\textrm{lg}_{\textrm{rot}}$, of each spot, obtained by:
\begin{equation}
\textrm{lg}_{\textrm{rot}} = \textrm{lg}_{\textrm{topo}} - 360^{\circ} \frac{nP_{\textrm{orb}}}{P_{\textrm{rot}}}.
\label{eq:rot}
\end{equation}
\noindent where the topocentric longitude, $\textrm{lg}_{\textrm{topo}}$, is the longitude in the fixed reference frame of the observer here on Earth, $n$ is the number of transits since the initial time of observation, $P_{\textrm{orb}}$ is the planet's orbital period, and $P_{\textrm{rot}}$ is the rotational period of the star at the transit latitude, where both periods are given in days. 

In the star's rotating reference frame, zero longitude is the projection of the line of sight during the central time of the first transit. Then, we vary the period of the star, $P_{\textrm{rot}}$, so that the spots become vertically aligned in a longitude map of the stellar surface at the transit band with time. This occurs because the same spot detected in a later transit should appear at the same longitude in the map of the stellar surface measured in the reference frame that rotates with the star.

To determine automatically the best rotation period, we first calculate the spot flux deficit, given by:
\begin{equation}
   \textrm{D}_{\textrm{flux}} = \sum r^{2}_{\textrm{spot}}(1-I_{\textrm{spot}}). 
   \label{eq:eqflux}
\end{equation}
\noindent where $r_{\textrm{spot}}$ and $I_{\textrm{spot}}$ are the radius and intensity of the spot, respectively.
The spot flux deficit is the sum of the flux of all spots  within a given longitude interval. This flux deficit is estimated
as a function of rotational longitude for each map with a given $P_{\textrm{rot}}$.
 
As the rotation period of the star, $P_{\textrm{rot}}$, is varied,  different functions of the spot flux deficit are produced. The auto-correlation function (ACF) of these flux deficits with the thinnest peak corresponds to the best rotation period of the star at that latitude band \citep{silva2011}. To determine the thinnest ACF, we calculate the Full Width Half Maximum (FWHM) of the main peak as a function of $P_{\textrm{rot}}$ and choose the smallest value. 

\subsection{Differential rotation}
A stellar rotation profile similar to that of the Sun is assumed for the stars in this study, which is given by:
\begin{equation}
    \Omega(\alpha) = A-B \sin^{2}(\alpha).
    \label{eq:omegaalpha}
\end{equation}
\noindent where $A$ and $B$ are the stellar equatorial angular velocity and rotational shear, respectively. Thus, the average angular rotation is given by:
\begin{equation}
\bar\Omega = \frac{1}{\alpha_{2}-\alpha_{1}} \int_{\alpha_{1}}^{\alpha_{2}} (A - B \sin^{2}\alpha)  \,d\alpha,
 \label{eq:meanomegalat}
\end{equation}
\noindent where $\alpha_{1}$ and $\alpha_{2}$ are the minimum and maximum latitude, respectively, where spots appear on the stellar disk. Assuming that $\alpha_{1} = 0^\circ$ and $\alpha_{2} = 90^\circ$, Eq.~\ref{eq:meanomegalat} reduces to:
\begin{equation}
\bar\Omega =  A - \frac{B} {2}. 
 \label{eq:meanomega}
\end{equation}

When only one planet is seen to transit its host star, the two unknowns of the rotation profile – the constants $A$ and $B$ - are estimated from two period measurements – the rotation period at the transit latitude, $2 \pi / \Omega(\alpha)$ (given by Eq.~\ref{eq:omegaalpha}), and the average rotation period from the maximum of the Lomb-Scargle periodogram, $\bar P = 2 \pi / \bar\Omega$ (given by Eq.~\ref{eq:meanomega}). 
Thus, solving Equations~\ref{eq:omegaalpha} and \ref{eq:meanomega} with the measured $\Omega(\alpha)$ and $\bar \Omega$,
the constants  $A$ and $B$ are determined, and hence the stellar differential rotational profile defined. The shear, or differential rotation, is given by $B = \Delta \Omega$ (rad d$^{-1}$) and the relative differential rotation by $\Delta \Omega / \bar \Omega$ (\%).

In the context of a planetary system, if there are multiple planets passing in front of the same star, it becomes possible to deduce several rotation periods occurring at various latitudes on the star's surface. These measurements of angular rotation at different latitudes, represented as $\Omega(\alpha_i)$, can be effectively modeled using Equation~\ref{eq:omegaalpha}, as exemplified in the Kepler-411 case.  Table~\ref{tab:stardif} presents the three distinct $\Delta\Omega$ values associated with each individual Kepler-411 planet, as well the overall result. More comprehensive details of these measurements are described in  \cite{araujo+21}.

\section{Results and Discussion}\label{sec:results}

\subsection{Differential rotation from spot transit mapping}\label{sec:keplerstars}

The analysis of stars differential rotation, from observations by the Kepler and CoRoT satellites, involved the application of the planetary transits mapping method. This method utilized longitude and latitude data from starspots as indicators of differential rotation. Table~\ref{tab:stardif} presents the parameters used to explore potential correlations, from left to right the columns list the star's name, effective temperature, average rotation period, and differential rotation $\Delta\Omega$ values, as well as the reference article.

The average rotation of the stars was estimated using Eq.~\ref{eq:meanomega} where the maximum latitude of spots was assumed to be $\alpha_2 = 90^\circ$. Since the previously published rotation shear of Kepler-71 was estimated considering $\alpha_2 = 65^\circ$ \citep{zaleski+19} and that of Kepler-210 for $\alpha_2 = 40^\circ$ \citep{valio22},  $\bar\Omega$ was recalculated with $\alpha_2 = 90^\circ$, and these  results are listed in Table~\ref{tab:stardif} for these two stars.
In the case of the Sun, its  white light curve from VIRGO observations over two solar cycles holds rotational information from all latitudes, not only the signatures of spots which occur up to 40$^\circ$ latitude. This occurs because not only spots add to the solar total brightness, but also bright faculae play an important part. 
Since faculae are observed at all latitudes, both surrounding active regions and as isolated features all the way to the poles \citep{Callebaut1992}, we considered all latitudes ($\alpha_2 = 90^\circ$) in the estimate of $\bar\Omega$. Nevertheless, we caution that the latitudinal distributions of spots in stars may not be the same as  that of the Sun over its cycle.

The uncertainties of  $\Delta\Omega$ listed in Table~\ref{tab:stardif} refer to the transit band width defined by the planet angular diameter. The nominal value refers to the shear calculated assuming the rotation period is that of the central latitude of the transit, where the upper and lower limits consider the latitudes of the transit band upper and lower edges.

\begin{table}
\setlength{\arrayrulewidth}{2\arrayrulewidth}
\centering
\small
\caption{Differential rotation for all stars.}
\setlength{\tabcolsep}{0.9mm}
\def\arraystretch{1.2}  
\begin{tabular}{lcccc}
\hline 
\hline
 Star &  T$_{\textrm{eff}}$ [K] &  $\bar P$ [d]& $\Delta\Omega$ [rad d$^{-1}$] &  Ref \\
 \hline
 \hline
Kepler-17    & 5781$\pm$85 & 12.4$\pm$0.1     & 0.0409$_{-0.0007}^{+0.0043}$ &  1 \\  
Kepler-45    & 3820$\pm$90 & 15.76$\pm$0.016    & 0.031$\pm$0.014            &  2, 3 \\ 
Kepler-71    & 5540$\pm$120 & 19.77$\pm$0.008    & 0.00098$\pm$0.00010       &  4,5 \\ 
Kepler-210   & 4559 & 12.33$\pm$0.002    & 0.044$^{+0.011}_{-0.007}$ &  6\\ 
Kepler-304   & 4731$\pm$100 & 14.18 $\pm$ 0.05   & 0.058$\pm$0.007 &  7,8\\
Kepler-411$^a$   & 4832 & 10.5 $\pm$ 0.03    & 0.0523 $\pm$ 0.0026  &  9 \\
Kepler-411$^b$   & 4832 & 10.5 $\pm$ 0.03    & 0.0514 $\pm$ 0.0017  &  9 \\
Kepler-411$^c$  & 4832 & 10.5 $\pm$ 0.03	    & 0.043 $\pm$ 0.019  &  9 \\
Kepler-411 & 4832 & 10.5 $\pm$ 0.03	    & 0.0500 $\pm$ 0.0006  &  9 \\
Kepler-1651  & 3594 & 18.43 $\pm$ 0.02    & 0.022$_{-0.003}^{+0.005}$  &  10  \\
CoRoT-2      & 5625$\pm$120 & 4.522$\pm$0.024     & 0.042$^{+0.005}_{-0.016}$  &  11,12 \\
CoRoT-4      & 6190$\pm$60& 8.90$\pm$1.1     & 0.0261$^{+0.0002}_{-0.0009}$&  11,12 \\ 
CoRoT-5      & 6100$\pm$65 & 26.63 $\pm$ 0.05   & 0.014$^{+0.001}_{-0.011}$   &  11,12\\ 
CoRoT-6      & 6090$\pm$50 & 6.40$\pm$0.05    & 0.105$^{+0.011}_{-0.08}$  &  11,12\\ 
CoRoT-8      & 5080$\pm$80 & 21.7 $\pm$ 0.5    & 0.014$^{+0.003}_{-0.008}$ & 11,12 \\ 
CoRoT-18     & 5440$\pm$100 & 5.4$\pm$0.40     & 0.094$\pm$ 0.003 & 11,12\\
\hline
\end{tabular}
\tablecomments{
(1) \citep{valio+17}; 
(2) \citep{zaleski2022dynamo};
(3) \citep{johnson2012characterizing};
(4) \citep{mathur2017revised};
(5) \citep{zaleski+19}; 
(6) \citep{valio22};
(7) \citep{zaleski2023};
(8) \citep{rowe+2014};
(9) \citep{araujo+21}; 
(10) \citep{1651};
(11) \citep{corot2023};
(12) \citep{bonomo2017gaps}.}

\label{tab:stardif}
\end{table}

\begin{figure}[H]
\centering
\includegraphics[scale=0.44]{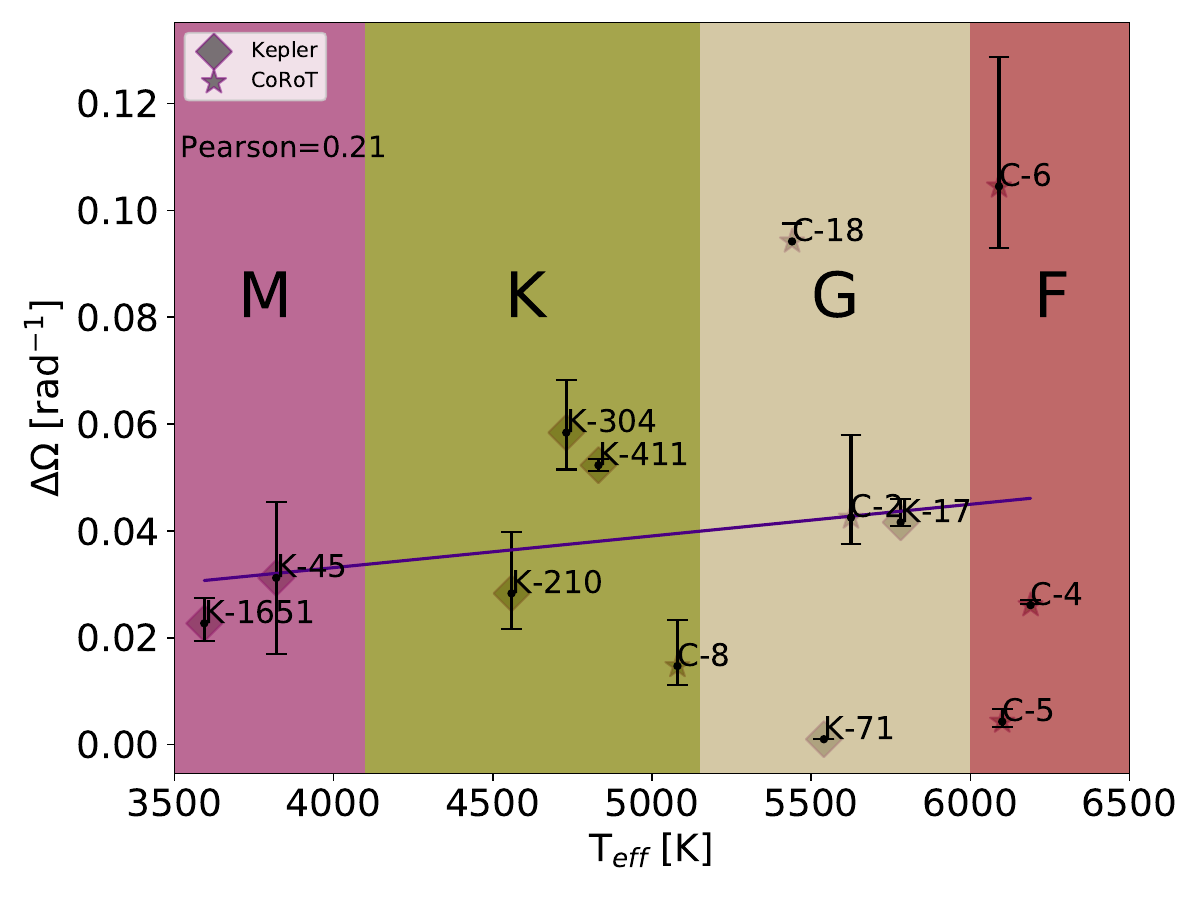}
\includegraphics[scale=0.44]{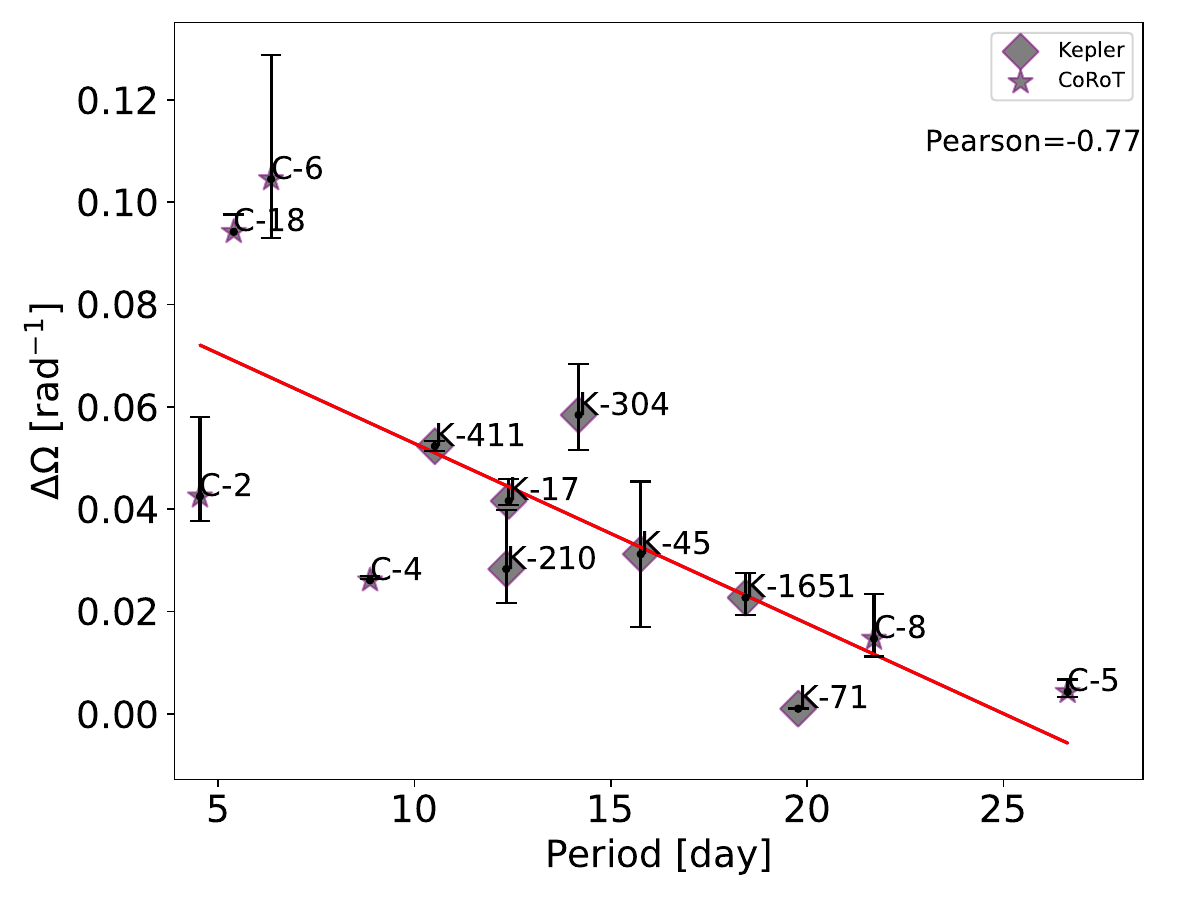}
\caption{Strong negative correlations for the dependence of $\Delta\Omega$ on the mean rotation period (\textbf{Right}) and a weak correlation for the dependence of $\Delta\Omega$ on the effective temperature of the stars (\textbf{Left}).}
\label{fig:rotation_temperature}
\end{figure}

The goal of our study is to investigate the relationship between  rotational shear, $\Delta\Omega$, and  stellar parameters such as  mean rotational period and photospheric temperature. The results are plotted in Figure~\ref{fig:rotation_temperature}, where the correlation between shear and rotation period is illustrated in the right panel, while the correlation between shear and $T_{\textrm{eff}}$ is shown in the left panel of Figure~\ref{fig:rotation_temperature}.

The results indicate a strong anti-correlation between  differential rotation with the average rotation period of  stars, yielding a Pearson correlation coefficient of -0.77. This suggests that as solar-like stars age or lose angular momentum, their rotation shear decreases.  However, we found a weak correlation between the rotational shear and effective temperature of the stars, with a Pearson correlation coefficient of 0.21, indicating that the shear does not depend on stellar mass. 

If we consider only the Kepler stars, the correlation index increases to -0.89 for the dependence of $\Delta\Omega$ on the average rotation period of the stars and drops to 0.036 for the $\Delta\Omega$ correlation with effective temperature.

We would like to point out that some of the stars analyzed show a commensurability between their rotation period and the planet orbital period, for example Kepler-17 and Kepler-71. This fact may be evidence of a star-planet interaction. \cite{Lanza2022a,Lanza2022b} studied the induction of resonant oscillations within the star's internal magnetic field due to a low-frequency component of the tidal potential due a close in planet. Despite the very week tidal potential, the oscillations are amplified due to the very long time interval of the star lifetime.
Nevertheless, these systems where there is a resonance between the periods of stellar rotation and planetary orbit do not behave differently in terms of their differential rotation, from the systems where the commensurability does not occur (see Figure~\ref{fig:rotation_temperature}).

The dependence of differential rotation has been investigated previously by several authors. For example,
\cite{barnes2005dependence} analyzed a total of 10 stars of spectral types G2-M2 using Zeeman Doppler imaging (ZDI) techniques, and  reported the existence of  a trend of surface differential rotation increasing with  effective temperature. Moreover, the authors found a power-law dependence of the horizontal shear $\Delta \Omega$  on the effective temperature given by $\Delta\Omega \propto T_{\textrm{eff}}^{8.92\pm 0.31}$. A similar trend was obtained from numerical modeling by \cite{kitchatinov2011differential}.
On the other hand,  more recently
\cite{reinhold2013rotation} observed that $\Delta \Omega$ weakly depends on the effective temperature of cool stars (3000–6000 K). For stars cooler than 6000 K, there is no notable trend of $\Delta \Omega$ with temperature \citep{reinhold2013rotation}. 

\cite{balona+16} present a review of the results available in the literature, where the rotation period was estimated for 2562 stars (K, G, F, and A). In their Figure 3, \cite{balona+16} show the results of the dependence of $\Delta \Omega$ as a function of rotation period and effective temperature, where $\Delta \Omega$ increases with effective temperature up to the convective boundary, but decreases slightly for  A stars.
In general, the authors agree that the shear increases with increasing rotation rate, that is, decrease with stellar rotation period. Furthermore, \cite{balona+16} explain that  the discrepancies which arise in the literature regarding the power-law dependence of $\Delta \Omega$ on effective temperature and rotation rate are most likely the result from an insufficient number of stars with known values of rotational shear. 

Studies have shown that there is a correlation between the magnetic activity of stars and their differential rotation, which is caused by the magnetic dynamo \citep{parker}. Young solar-type stars are known to have faster rotation periods and stronger magnetic activity compared to the Sun \citep{frohlich2012magnetic}. This relationship is evident in other phenomena associated with the stellar dynamo, including the presence of large stellar spots, such as those observed in Kepler-17 with mean radii of 49,000 $\pm$ 10,000 km \citep{valio+17}, and the production of superflares and stellar spots with complex magnetic topology \citep{araujo2023connection}. 

\section{Conclusion}\label{sec:conclusion}

We have used the spot transit mapping method to investigate the relationship between stellar differential rotation with the mean rotation period and effective temperature of FGK and M stars. Albeit being limited to stars with transiting planets, we believe this method to be a precise estimate of differential rotation given the higher spacial resolution of spots signature. Moreover, the transit mapping method can also be applied to slowly rotating stars, which are not suitable for ZDI technique analyses.

The differential rotation of 13 stars of spectral types M through F, observed by the Kepler and CoRoT satellites, were analyzed using the spot transit mapping technique. Correlations between rotational shear and the rotation period and effective temperature of these stars were sought. We found no significant correlation between the rotational shear and stellar effective temperature, however, a strong anti-correlation between the rotational shear and the average stellar rotation period was observed. We point out that all stars studied here were slow rotators ($\bar P \ge 5$ days).

One important characteristic of these solar-type stars is that they are more active than the Sun, with more numerous and larger spots. Notably, the dynamos operating in highly active stars, particularly spectral type G stars, are likely to be significantly distinct from that of the Sun. \cite{donati2003temporal} suggest that dynamo processes in these stars are likely distributed throughout the convective envelope, instead of being confined at its base as in the Sun. This finding may explain the dispersion of rotational shear with effective temperature observed in these stars.

A clear understanding of the relationship between rotational shear with the effective temperature and rotation rate of a star is necessary to advance our knowledge of stellar dynamos. Moreover, the different behaviour of fast and slow rotators implies that age also plays a key role in differential rotation.
All this information is crucial for understanding the behaviour and evolution of stars, including the Sun, and for improving our prediction of their behaviour.

\section*{Acknowledgements}

We are grateful to the anonymous referee for the valuable suggestions that improved this work. 
Furthermore, we acknowledge the partial financial support received from FAPESP grants \#2013/10559-5, \#2018/04055-8, and \#2021/02120-0, CNPq grant \#150817/2022-3, as well as MackPesquisa funding agency.

\bibliography{sample631}{}
\bibliographystyle{aasjournal}

\end{document}